\documentclass[10pt, conference, compsocconf]{IEEEtran}
\IEEEoverridecommandlockouts

\usepackage{graphicx}
\usepackage{epstopdf}
\usepackage{array}
\usepackage{tabu}

\usepackage[yyyymmdd,hhmmss]{datetime}

\usepackage{blindtext}

\usepackage[usenames]{color}   

\renewcommand{\baselinestretch}{.86}

\newcommand{\hide}[1]{}   

\begin{document}

\title{
\vspace*{-2cm}
\fbox{\small Preprint from Proc. IEEE Big Data Congress, July, 2015} \\
\vspace*{1cm}
Alexandria: Extensible Framework for Rapid Exploration of Social 
Media$^\dag$\thanks{$^\dag$ This material is based upon work supported by the U.S. Defense Advanced Research Projects Agency (DARPA) under Agreement Number W911NF-12-C-0028. The content of the information in this document does not necessarily reflect the position or the policy of the Government, and no official endorsement should be inferred. The U.S. Government is authorized to reproduce and distribute reprints for Government purposes notwithstanding any copyright notation here on.}}

\author{\IEEEauthorblockN{Fenno F. Heath III, Richard Hull, Elham Khabiri, \\ 
Matthew Riemer, Noi Sukaviriya, and  Roman Vacul{\'{\i}}n}
\IEEEauthorblockA{IBM Research \\
Yorktown Heights, New York, USA \\
Email: \{theath, hull, ekhabiri, mdriemer, noi, vaculin\}@us.ibm.com}
}





%
%

\maketitle


\begin{abstract}
The Alexandria system under development at IBM Research provides an extensible framework and platform for supporting a variety of big-data analytics and visualizations. The system is currently focused on enabling rapid exploration of text-based social media data. The system provides tools to help with constructing ``domain models'' (i.e., families of keywords and extractors to enable focus on tweets and other social media documents relevant to a project), to rapidly extract and segment the relevant social media and its authors, to apply further analytics (such as finding trends and anomalous terms), and visualizing the results. The system architecture is centered around a variety of REST-based service APIs to enable flexible orchestration of the system capabilities; these are especially useful to support knowledge-worker driven iterative exploration of social phenomena. The architecture also enables rapid integration of Alexandria capabilities with other social media analytics system, as has been demonstrated through an integration with IBM Research's SystemG. This paper describes a prototypical usage scenario for Alexandria, along with the architecture and key underlying analytics. 
\end{abstract}

\section{Introduction}
\label{sec:intro}

Twitter, Instagram, forums, blogs, on-line debates, and many other
forms of social media have become the outlets for people to freely and
frequently express ideas.  Indeed, many research papers
have explored 
social media usage in many application
areas.
Research has ranged from using social science techniques to
find indicators of phenomena such as increased health risks,
to studies on optimization of hugely scaled analytics computations,
to usability of analytics visualizations.
However, there has been little work on how to
bring together the myriad of analytics capabilities to
support knowledgable business analysts 
in
rapid, collaborative, and iterative exploration and analysis
of large data sets.
This requires a combination of several aspects,
including integration of numerous analytics tools,
efficient and scalable data and processing management,
a unified approach for data and results visualization,
and strong support for on-going knowledge-worker driven activity
to uncover and focus in on particular areas of interest.
The paper describes the Alexandria system, currently 
under development at IBM Research, which supports these
several aspects.
The system is currently focused on the early stages of
the overall analytics lifecycle, namely, on enabling
rapid, iterative exploration and visualization of social media data in connection with
a given domain (e.g., consumption habits for beverages, 
the growth of the market for vegan foods, or political opinions
about an upcoming election).
The system has been designed to support rich extensibility,
and has already been integrated with a complimentary system
at IBM.

\begin{figure} 
\vspace*{-2mm}
\centerline{\includegraphics[width=3.5in]{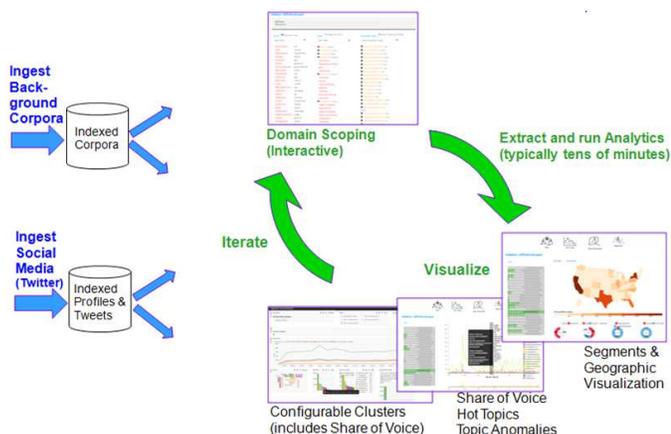}}
\vspace*{-3mm}
\caption{Alexandria supports iterative exploration of social media, 
and includes background text analytics processing. 
(See also Figure \ref{fig:arch-flow}.)}
\label{fig:Alex-iterative-usage}
\vspace*{-2mm}
\end{figure}

Figure \ref{fig:Alex-iterative-usage} shows the two main parts of
current Alexandria
processing, namely Background Processing and Iterative Exploration.
The Background Processing includes primarily (a) various analytics on
background text corpora that support several functionalities,
including similar term generation, parts-of-speech and collocation
analytics, and term-frequency-inverse-document-frequency (TF-IDF) 
analytics; and (b) ingestion and indexing of social media data 
(currently from Twitter) to enable main-memory access speeds
against both text and structured document attributes.
(Although not shown in the figure, there is also background
analytics to compute selected author profile attributes, e.g., 
geographic location, family aspects, interests).
Iterative Exploration enables users to build a number
of related {\em Projects} as part of an investigation of some 
domain of interest.
Each Project includes 
(i) the creation of a targeted {\em domain model}
used to focus on families of tweets and authors relevant to the investigation,
(ii) application of a variety of analytics against the
selected tweets and their authors, 
and
(iii) several interactive visualizations of the resulting analtyics.
At the beginning of an investigation there are
typically several {\em experimental} Projects, used by
individuals or small collaborating groups.
Over time some Projects
may be {\em published} with more stability for broader usage.

Alexandria advances the state of the art of
social media analytics in two fundamental ways
(see also Section \ref{sec:related}).
First, the system brings together several text analytics tools
to provide a broad-based environment to rapidly
create domain models.  
This contrasts with research that has focused on perfecting
such tools in isolation.
Second, Alexandria applies data-centric and other design principles
to provide a working platform that supports 
ad hoc, iterative, and collaborative exploration of social media data.
As such, the system extends upon themes presented
in \cite{rajaraman2011mining,analytics-process-mgmt:DAB-2014}, 
and develops them in
the context of social media applications.

Section \ref{sec:goals} highlights the key goals for Alexandria, including
both longer- and shorter-term ones.
Section \ref{sec:overview} describes a prototypical usage scenario for
the system, and illustrates its key functionalities.
Section \ref{sec:arch} highlights key aspects of the system architecture,
and describes how the design choices support the key goals.
Section \ref{sec:scoping} describes key technology underpinnings
for the domain scoping capability, and
Section \ref{sec:analytics} does the same for the currently supported analytics.
Section \ref{sec:meta-data} describes the data-centric approach taken
for managing exploratory Projects to enable rich flexibility.
Section \ref{sec:related} describes related work,
and Section \ref{sec:conc} discusses future directions.

\section{System Goals}
\label{sec:goals}

This section outlines the primary long- and shorter-term goals that have
motivated the design of the Alexandria framework and system.

\hide{

The longer-term goal is to provide an extensible approach
to support a broad variety of analytics, with 
an eye towards supporting the overall analytics
lifecycle -- from exploration to callibration to production usage.
A shorter-term goal is to provide a flexible, easy to use, extensible
system for rapid exploration of social media.

} 

The longer-term goals are as follows:
\smallskip

\noindent
{\bf LG1: Extensible platform to support business users with
numerous styles of analytics.}
This contrasts significantly with most previous works,
that are focused primarily on scalable performance,
support for targeted application areas,
or support primarily for data scientists. 
\hide{
While there have been many advances in
recent years in tools for so-called Big Data Analytics,
most of the work has focused on
the development of new analytics algorithms,
improving the speed of widely used algorithms,
the development of highly scalable general-purpose
platforms (e.g., SPARK),
or the development of platforms for selected
kinds of data (e.g., TITAN for graph data).
There are also tools to support the data mining
process itself (e.g., IBM's SPSS, or the opensource RapidMiner).
} 
Alexandria is focused on providing a layer above all of these,
to enable business users to more effectively use analytics, both
to find actionable insights, 
and also to incorporate them 
into on-going business processes.

\hide{ 

As such, Alexandria strives to expose different analytics 
capabilities, and give users rapid access to them.
Implementation and optimization details are largely
``under the hood''. 
Configuration of parameters should
be largely defaulted, and optionally adjustable by
more sophisticated users.

} 

\hide{

The system should provide extensibility in two main
directions. 
First, it is important to support new kinds of data and
new kinds of analytics as they become available or relevant.
Second,
because the technology for analytics processing
and visualization continues to advance, and because
different systems are more appropriate for different kinds of
data,
it is important
to provide an abstraction layer above the particular implementation
approaches.

} 

\hide {

While REST services APIs are a natural choice to
enable a uniform approach for enabling the integration of
numerous data sources, analytics, and visualizations,
mechanisms must also be provided to pass data ``by reference'',
and indeed, to enable data to remain ``in place'' as much as 
possible through multiple analytics steps. 

} 

\smallskip

\noindent
{\bf LG2: Support analytics process lifecycle, from exploration to prescription.}
As discussed in 
\cite{analytics-process-mgmt:DAB-2014},
there are several stages in the lifecycle of analytics usage,
ranging from initial exploration, to refinement and hardening,
to incorporation into already existing business processes for
continuing value add, 
to expanding the application to additional aspects of a business.
While the CRISP-DM method \cite{CRISP-DM}
addresses several elements of the lifecycle, 
the method and associated tools are
geared towards data scientists rather than business users.
In contrast, a goal of Alexandria is to provide
business users with substantial exploration capabilities,
and also support the evolution of analytics approaches from
exploration to production usage.
Of course data scientists will still play a very key role, 
and the Alexandria platform should
enable graceful incorporation of new algorithms as they
become available from the data scientists.

\hide{

On a more concrete level, 
Alexandria is intended to support the following
key analytics lifecycle capabilities.
  \begin{enumerate}
  \item Make it easy to explore and 
        understand large volumes of unstructured data
  \item Enable non-subject matter experts to explore a
        domain of interest, find relevant social media, and discover insights
  \item Support an interactive experience, including in particular
        fast response times wherever possible
  \item Enable flexible, composable intuitive analytic components
  \item Help users find trends and emergent phenomena
  \end{enumerate}

} 


\noindent
{\bf LG3: Support for a collaborative production environment.}
Analytics is no longer the realm of a small team of data scientists
working largely in isolation.  
Rather, it is increasingly performed by a multi-disciplinary
team that is in parallel digging more deeply into the data,
finding ways to add business value by
integrating analytics insights into 
existing business processes,
and finding ways to make the usage of the insights 
production grade.
\medskip

\hide{

As such, the Alexandria platform should support
several non-functional capabilities, including the following. 
  \begin{enumerate}
  \item Enable sharable results, including catalogs of both building-block
        and self-contained analytics
  \item Enable repeatability and cloning of analytics flows
        and Projects, as enabled through rich mechanisms for
        recording and accessing provenance information
  \item Support multiple levels of multi-tenancy, as well
        as easy configurability of existing analytics to
        block access to components that given users groups
        may not have licenses or authorization to use
  \end{enumerate}

} 

\noindent
{\bf LG4: Scalable, e.g., work with billions of tweets and forum comments.}
The Alexandria system should be able to work with
state-of-the-art systems such as SPARK and TITAN, and
more generally with Hadoop-based and other distributed
data processing systems, to enable rapid turn-around
on large analytics processes. 
Similarly, the system should
support main-memory indexing systems such as
Elastic Search or LUCENE/SOLR to enable split-second
access from very large data sets, including text-based searches.
\medskip

As a way to get started with the longer-term goals,
the initial version of Alexandria has focused more
narrowly on (a) Social Media analytics, and on (b) the
exploration and initial visualization phases of the
overall analytics process.
The key shorter-term goals include the following:

\def\LGone{{\bf LG1}}
\def\LGtwo{{\bf LG2}}
\def\LGthree{{\bf LG3}}
\def\LGfour{{\bf LG4}}

\def\SGone{{\bf SG1}}
\def\SGtwo{{\bf SG2}}
\def\SGthree{{\bf SG3}}
\def\SGfour{{\bf SG4}}
\def\SGfive{{\bf SG5}}
\def\SGsix{{\bf SG6}}

\noindent
\SGone:
Enable users to begin their exploration of a new topic
domain within a matter of hours.
\smallskip

\noindent
\SGtwo:
In particular, enable non-experts to quickly create a
domain model (i.e., keywords and extractors) that enables
a focus on Tweets and other social media that 
are relevant to a given topic.
\smallskip

\noindent
\SGthree:
Provide a variety of different analytics-produced views of the data,
to permit different styles of data and results examination
\smallskip

\noindent
\SGfour:
Support iterative exploration based on info learned so far, 
including management of meta-data about raw and derived data sets
\smallskip

\noindent
\SGfive:
Minimize processing time through
to enable as much interactivity as
possible, by using main-memory indexes,
parallel processing, avoiding data transfers, etc.
\smallskip

\noindent
\SGsix:
Enable easy and fast orchestration of capabilities, including
rapid creation of variations on the domain model and the
analytics processing.  
This includes the automation of processing steps and
the defaulting of configuration parameters 
wherever possible.

\section{Using the System}
\label{sec:overview}

\hide{

\begin{figure} 
\vspace*{-2mm}
\centerline{\includegraphics[width=3.5in]{Alex-overall-flow.eps}}
\vspace*{-3mm}
\caption{Key functional components of current Alexandria system.  (Blue
indicates background processing and green indicates 
user-requested processing}
\label{fig:Alex-overall-flow}
\end{figure}

} 

This section illustrates the main capabiliites currently supported in
Alexandria through an extended example.

\hide{ 

Figure \ref{fig:Alex-overall-flow} indicates the major processing steps of the
system; the user is directly involved with the five stages of activity shown in green,
and also the orchestration of those stages.
These are described in the following.

} 

To extract “relevant” documents from social media, one needs to gather documents that mentioned terms, expressions, or opinions pertaining to the area one wants to explore.  
Alexandria provides tools that support both laymen and experts in
finding terms that cover the space of interest, and also terms that can
drill more deeply into that space.

\hide{ 
Challenges are (1) we may not know what “good” relevant terms or topics are, especially in the absence of knowledge in the area.  (2) There are many ways to express one same thing hence it’s hard to know how to catch them all.  This is often true with or without the knowledge of the area of exploration.
Alexandria is designed and implemented to alleviate the burden of these two challenges.  This is evident in the first initial two steps when using the system.  In Alexandria, once an area of exploration is identified, the discovery process is an iterative cycle between scoping the area – expanding the breadth by finding more topics related to the area – and focusing – ensuring the depth by accounting for variations of the chosen topics.  This section will demonstrate how Alexandria supports this process.
} 

We will explore a subject around vaccination as an example for this paper.  
Suppose that the government would like to encourage people to take vaccination, but wonder what people’s opinions may be around vaccination.  The exploration starts with creating a Project with a few seed terms,
namely `vaccination', `flu' and `measles'.  Based on these terms, we asked Alexandria to generate a family
of relevant collocated 
terms in an effort to bound the scope.
These terms may be manually edited, to reach the terms listed in Figure \ref{fig:termGen-with-edits}.
Here, the black terms were generated automatically,
red were added by hand, and gray with strike out were 
auto-generated but deleted by hand.

\hide{
The user might be someone who is knowledgeable about the area being explored, or might be interested but lacks the vocabulary to provide breath to the exploration.   Alexandria can be useful for both types of users.  Those with familiarity with the subject may use Alexandria as a quick bootstrap where they can provide additional terms to explore after some terms are generated.  Those without much of the subject knowledge will use Alexandria to gain breadth that oth-erwise would not be easily obtained.
}

In some cases the auto-generated terms will help the user learn more about the domain of
interest.
In this example, Dr. Anne arises, and a Google search reveals that Dr. Anne Schuchat is
the Director of the U.S. Center for Disease Control \cite{anne-schuchat-web-site}, 
so her name was left in the list.
Similarly, Dr. Gil remains because he is mentioned in a news article \cite{measles-in-disneyland}
concerning a measles outbreak at Disneryland.

\hide{
Please note in Figure \ref{fig:termGen-with-edits} that Alexandria’s generated terms may not be all terms the users need.  On one hand, the user can prune these terms as they see fit for various reasons.  Those that are not relevant or consi-dered out of scope can be removed.  Some of the terms are removed mainly because they do not make sense in the context of social media search.  On the other hand, new terms can be added and existing terms modified to shape the scope that fits their needs.  Some of the terms require some research such as Dr. Anne as it may lead to something interesting.  We used Google with the key words Dr. Anne and vaccination to find out Dr.AnneSchuchat is the Director of Central of Disease Control, hence we left her name in the list.  Her last name was included later in the list.  We also left Dr. Gil in the list as he was mentioned in a news article concerning a measles outbreak at Disneryland.  We didn’t add any terms in this example. Table 2 shows the list from Table 1 pruned and enhanced.  Terms in color orange are terms modified by us

} 

\hide{

\begin{figure}
\begin{center}
\vspace*{-2mm}
\centerline{\includegraphics[width=3.5in]{termGen-no-edits.eps}}
\vspace*{-5mm}
\end{center}
\label{fig:termGen-no-edits}
\caption{List of relevant terms collocated with the seed terms used to launch Project}
\end{figure}

} 

While the scoping step is supposed to extend our vocabulary to cover various areas of the topics, some terms appear to be rather similar.  For example, many variations of vaccination are included in the list.  We know that if a tweet mentioned one of these terms, it is likely to have something to do with vaccination.  Alexandria supports automatically clustering similar terms into groups called “topics.”  Each topic is used to provide a list that, if a tweet mentions one or more of the terms in the topic, we can conclude that the tweet has mentioning of this topic.

\begin{figure}
\begin{center}
\vspace*{-2mm}
\centerline{\includegraphics[width=3.5in]{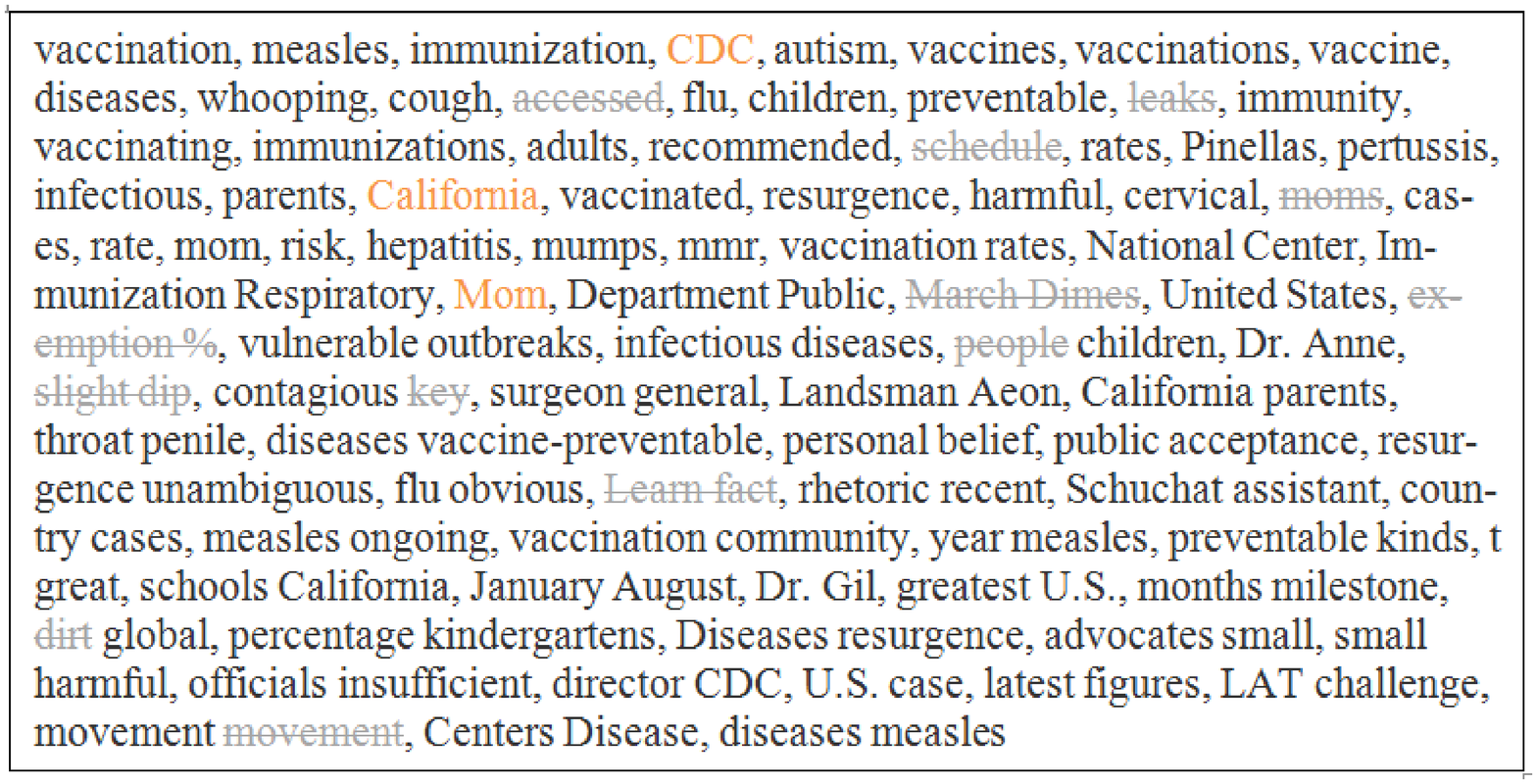}}
\vspace*{-3mm}
\end{center}
\label{fig:termGen-with-edits}
\caption{List of relevant terms collocated with the seed terms, after manual edits}
\end{figure}

\begin{figure} 
\vspace*{-2mm}
\centerline{\includegraphics[width=3.5in]{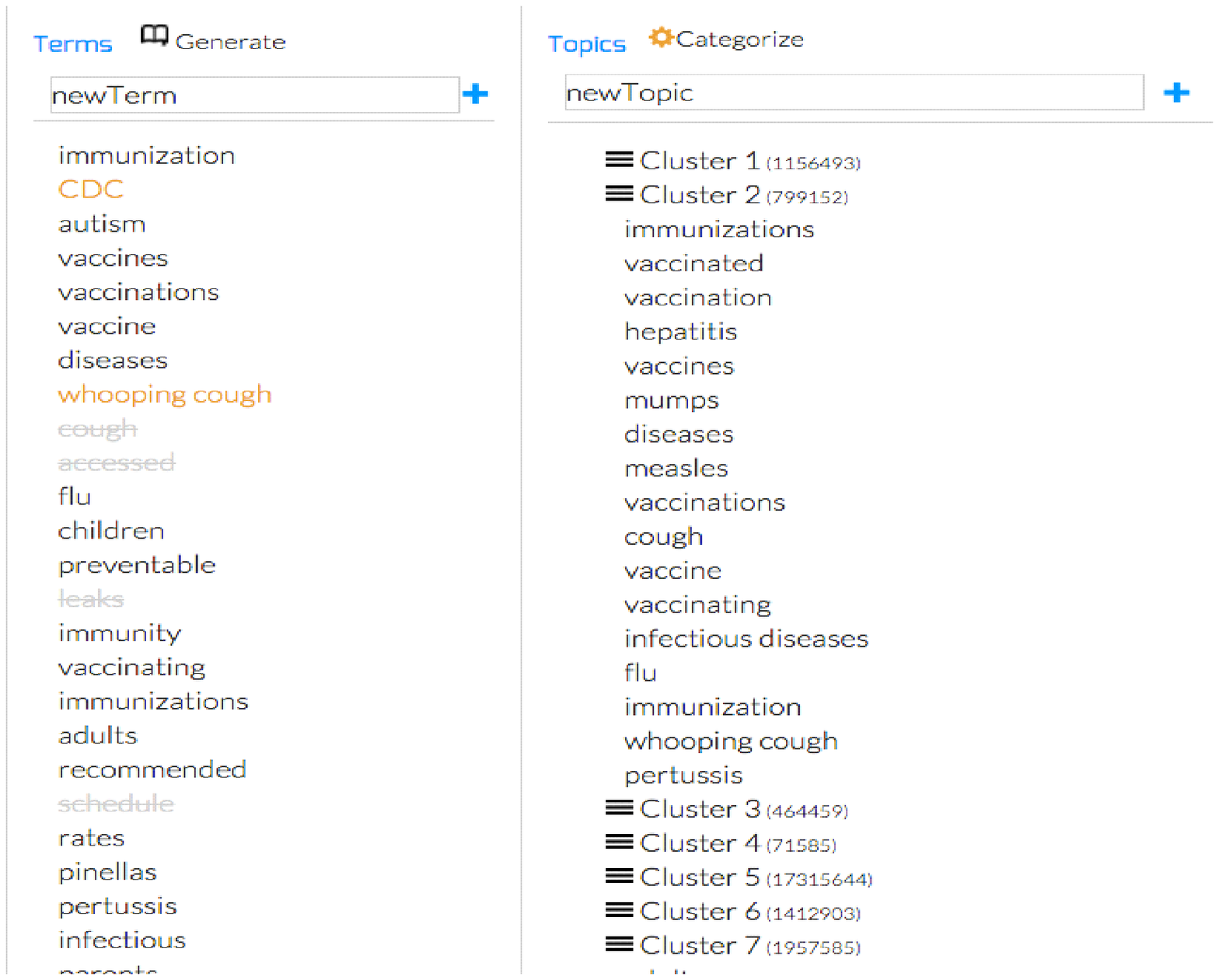}} 
\vspace*{-3mm}
\caption{Alexandria interface for domain scoping: After automatic term clustering}
\label{fig:scoping-01-terms-open-concepts}
\vspace*{-2mm}
\end{figure}

Figure \ref{fig:scoping-01-terms-open-concepts} above shows a snippet from the actual Alexandria page where the terms are listed vertically in the first column and the second column shows the clusters suggested by Alexandria.  Note that these clusters are generically named “Cluster 1,” “Cluster 2,” and so forth.  In the figure
Cluster 2 is ``open'', to show terms Alexandria placed in it.  It appears many vaccination and diseases that can be prevented by vaccination are included.  In Section \ref{sec:scoping}, we detail the analytics we are doing behind the scenes for topic clustering.  

\hide{

\begin{figure} 
\vspace*{-2mm}
\centerline{\includegraphics[width=3.5in]{scoping-02-concepts-after-edits.eps}}
\vspace*{-3mm}
\caption{Topics after organizing around opinions about vaccinations}
\label{fig:scoping-02-concepts-after-edits}
\end{figure}

}

\begin{figure} 
\vspace*{-2mm}
\centerline{\includegraphics[width=3.5in]{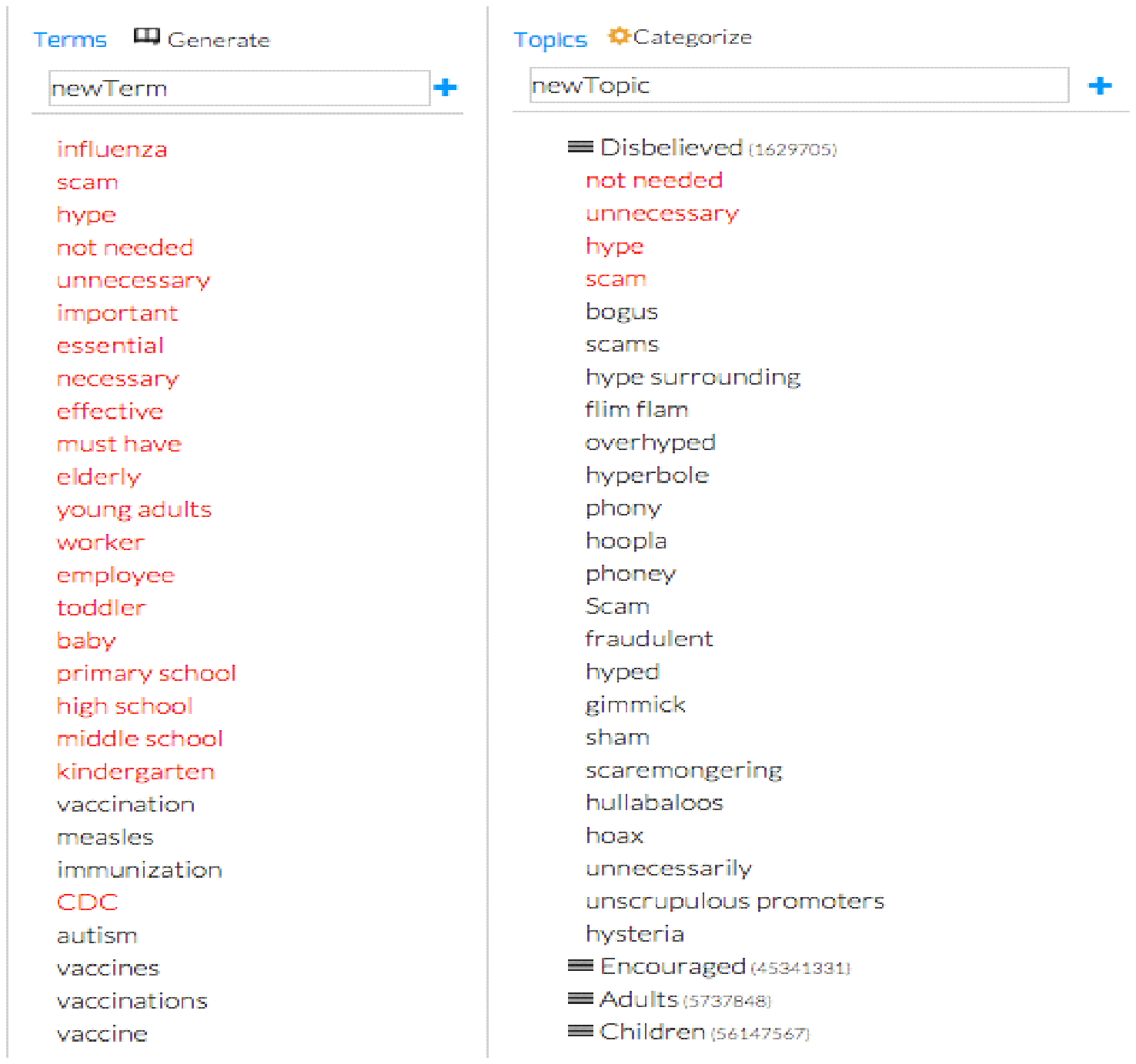}} 
\vspace*{-3mm}
\caption{Topics after adding terms in the ``Disbelieved'' topic}
\label{fig:scoping-03-adding-disbelieved}
\vspace*{-2mm}
\end{figure}

The numbers to the right of each topic indicate the number of tweets in which at least one term from the topic is found.  One can use this number to gauge how widespread the topic is.  Bear in mind that something general such as “Disbelieved” can be about any subject, hence the large number of over a million tweets, and not necessarily about vaccination.  
These numbers are obtained within seconds through accesses to a SOLR index holding all of the tweet information.

\begin{figure} 
\vspace*{-2mm}
\centerline{\includegraphics[width=3.5in]{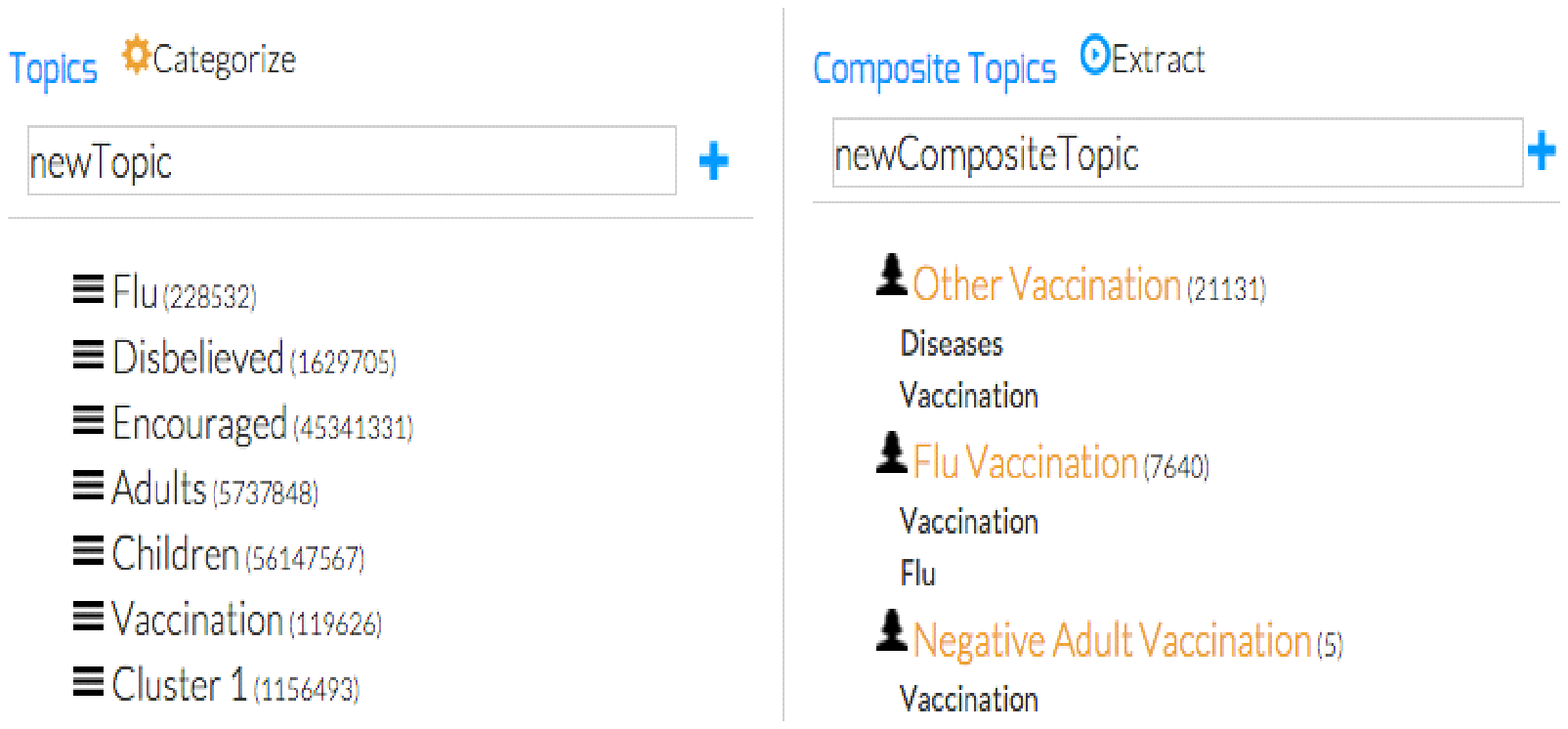}} 
\vspace*{-3mm}
\caption{Topics after adding terms in the ``Disbelieved'' topic}
\label{fig:scoping-04-composite-topics}
\vspace*{-2mm}
\end{figure}

Figure \ref{fig:scoping-04-composite-topics} illustrates the state of the system after a few steps.
First, Alexandria supports renaming of the clusters, and moving them around in the middle column.
Second, there is an automated ``similar term generation'' service for adding depth to an existing topic.
In the figure, the red terms in Disbelieved where inserted by hand, and the
terms in black below that were generated automatically to add depth.

\hide{

Once the topics are settled as shown in Figure 2, we requested Alexandria to do exactly that.  Figure 3 shows terms in the group “Disbelieved,” which now has many more added to the topic to the original 4 we created.  Notice also the number of tweets for “Disbelieved” has increased to over 1.6 million.  As a matter of facts, the numbers of tweets for all topics have increased after depth of topics is added, showing that by increasing the depth of each topic, we have a better reach in social media exploration.   Because of space limitation for this paper, we will not show additional terms in other topics.

} 

The third phase of scoping is the
building of the actual extractors (or queries) for selecting tweets of interest.
This is accomplished by creating
“composite topics,” which are based on Boolean combinations of the topics.
Figure \ref{fig:scoping-04-composite-topics} shows several composite topics,
some of which are ``open'' to expose the topics that are used to form them.
(At present the UI supports only conjunctions of topics, but the underlying engine
supports arbitrary combinations.)
For example, a composite topic “Support Flu Vaccination” we combine “Flu,” “Vaccination” and “Encouraged” topics to form a search statement of “find any tweets that mentioned at least one (or more) of the terms in “Flu” and one (or more) of the terms in “Vaccination” and one (or more) of the terms in “Encouraged.”  
(A further refinement would be to exclude tweets that include a negating term such as ``not''.)


Once the set of composite topics has been specified, it
is time to perform some data extractions, re-structurings, and indexing to
support various anlaytics.
Upon request, Alexandria extracts tweets with topics matching the composite topic combinations, annotates each tweet accordingly, and then launches multiple analytics activities on these tweets.  One of the activities was extracting the author profiles of these tweets and aggregate attributes among these profiles.  We will detail this work on in Section \ref{sec:analytics} on Analytic View.


\hide{
\begin{figure*}[t] 
\vspace*{-2mm}
\centerline{\includegraphics[width=6.5in]{FIG/scoping-05-terms-concepts-indicators}} 
\vspace*{-3mm}
\caption{Alexandria interface for domain scoping: After creating composite topics}
\label{fig:scoping-05-terms-concepts-indicators}
\end{figure*}
} 

\begin{figure} 
\vspace*{-2mm}
\centerline{\includegraphics[width=3.5in]{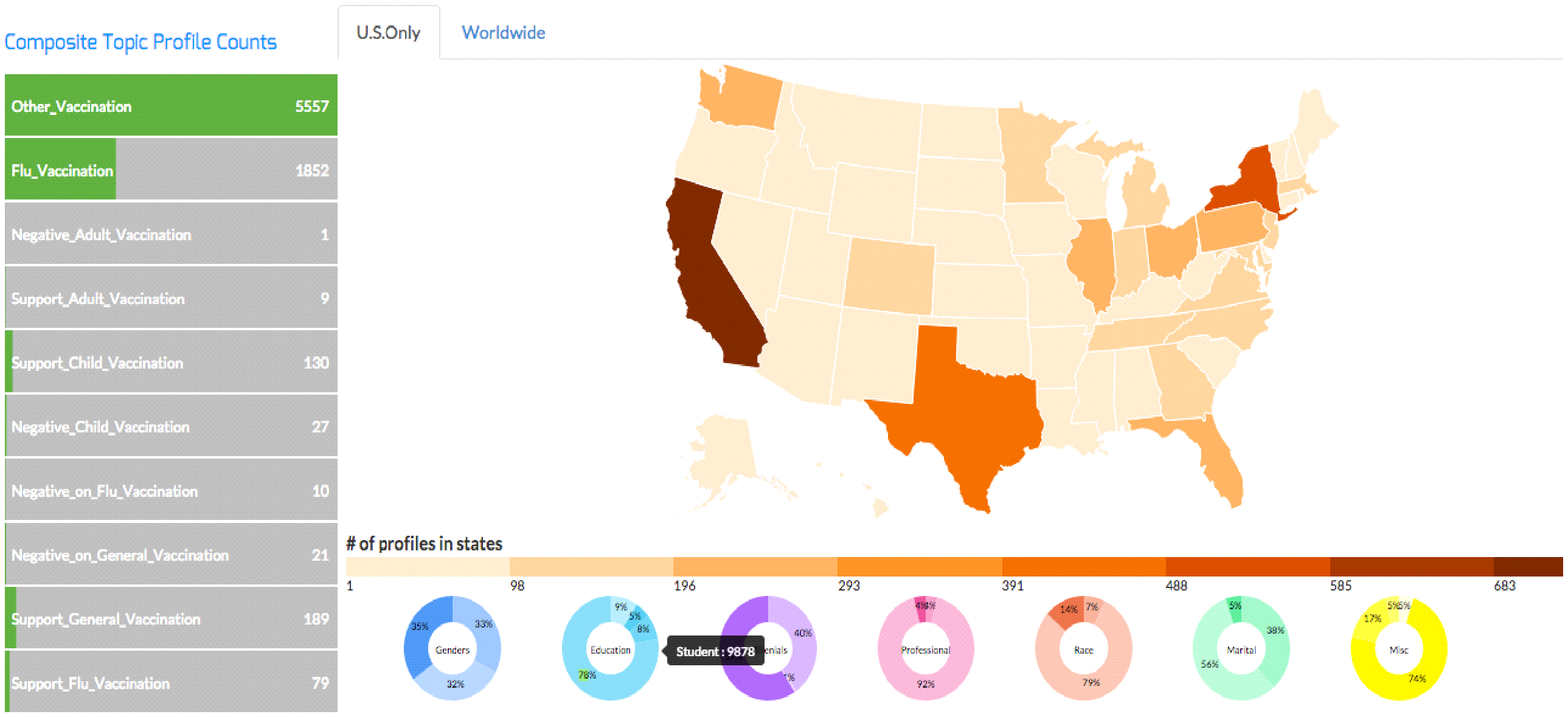}} 
\vspace*{-3mm}
\caption{Interactive view for exploring the demographic distribution of tweet authors 
          who are negative about flu vaccination}

\label{fig:views-map}
\vspace*{-2mm}
\end{figure}

We now describe some of the visualizations used to show the analytics
results associated with a Project.  
In one direction, Alexandria infers profile attributes of Twitter
authors through background analysis of 100's tweets per author.
Information such
as education, gender, ethnicity, location of residence is 
inferred based on evidence of words found in tweets.  Figure
\ref{fig:views-map} shows how the demographic distribution of tweet
authors of composite topics in the U.S.  On the left, it shows the
numbers of authors for various composite topics.  On the map, states
with darker colors mean higher numbers of authors reside in those
states.  Mousing over a state (not shown) would give more details of
these authors.  The colored donuts below the map show percentage of
various characteristicsof those located in the U.S. for example, male,
female or unknown for gender.  Mousing over a portion of a donut shows
the value of the characteristics and the number of profiles.  For
example, in the figure we show that 5898 tweet authors of all topics
combined are students.

\begin{figure} 
\vspace*{-2mm}
\centerline{\includegraphics[width=3.5in]{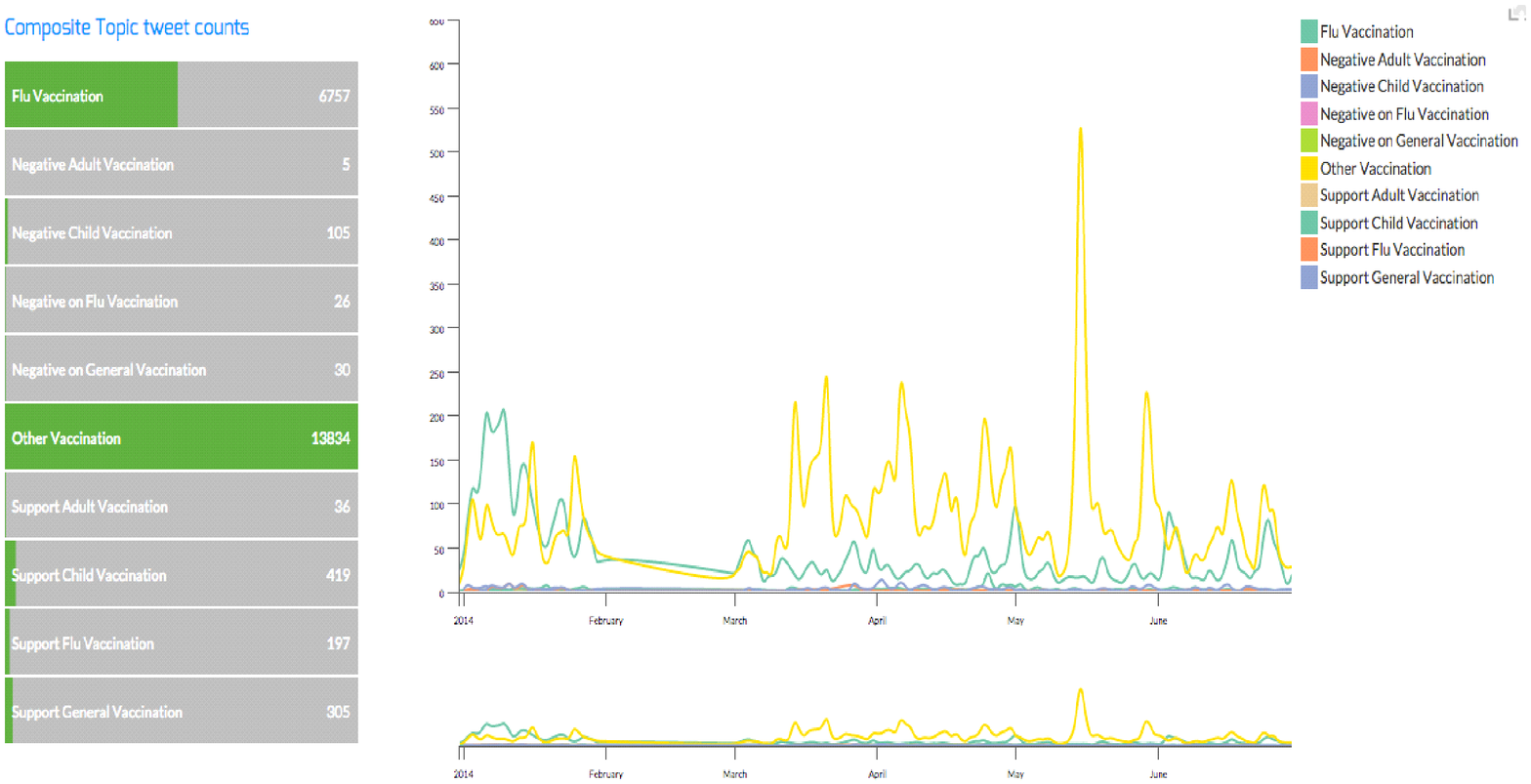}} 
\vspace*{-3mm}
\caption{Share of voices of tweets from different composite topics}
\label{fig:views-topic-anomaly}
\vspace*{-1mm}
\end{figure}

\begin{figure} 
\vspace*{-2mm}
\centerline{\includegraphics[width=3.5in]{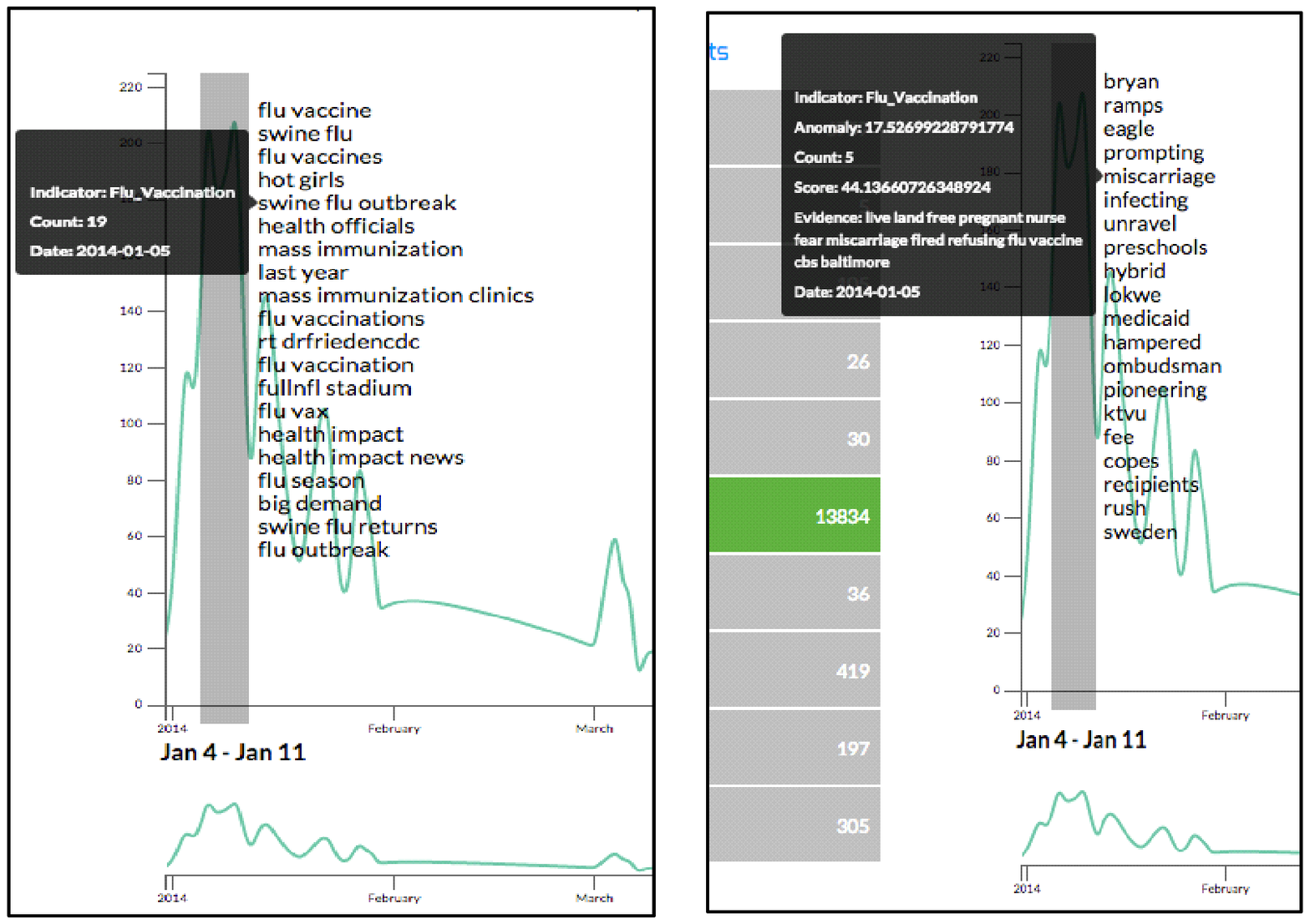}} 
\vspace*{-3mm}
\caption{Exploring frequently mentioned and anomalous terms for the composite topic “Flu Vaccination” in the week of Jan4 to Jan 11.   The text boxes show contextual data of the term they point to.  The box on the right also shows partialcontext of the tweets where the term was extracted from.}
\label{fig:views-black-boxes}
\vspace*{-2mm}
\end{figure}

Figure \ref{fig:views-topic-anomaly} 
illustrates another analytic view in which Alexandria shows “share of voices”,
i.e., comparison of tweet volumes of the composite topics over time.  In this paper we are working on tweets from January to June of 2014.  
Notice the higher volumes among the topics “Flu vaccination” and “Other Vaccination” in 
Figure \ref{fig:views-topic-anomaly}, with a peak around mid-May for “Other vaccination” topic.  One may wonder what happened during that week.  In this view we can click on the graph to explore the frequently mentioned terms or anomalous terms mentioned in that week.  Figure \ref{fig:views-black-boxes} shows  
snippets of two images captured to highlight the two types of terms.

Specifically for Figure \ref{fig:views-black-boxes}, we selected the
“Flu Vaccination” topic on the left to narrow the visualization down
to just this topic, hence the presence of only one line graph in the two
snippets.  This line represents the volume over time of tweets
that match the “Flu Vaccination” extractor.  For this topic, there
seems to be a peak around the second week in January. The snippet on
the left of the figure shows frequently mentioned terms in the week
while the snippet on the right shows terms that are considered
anomalous in that week.  We moused over the term “swine flu outbreak”
which was mentioned 19 times, hence showing up high in the word list.
However, this term is not considered anomalous, indicating that
this term also shows up fairly often in other weeks.
However, the term “miscarriage” is anomalous.  Mousing over it reveals
some evidence of the news about a nurse refusing to get vaccinated and
subsequently being fired from a hospital.

There are other views that one can use to explore analytics results of social media insights,
including some that leverage the configurable Banana visualization tool \cite{banana}.  


\hide{
\begin{figure}[t]
\center{\includegraphics[width=3.5in]
        {FIG/lifecycle.png}}
\caption{The lifecycle model of {\tt CustomerOrder}}
\label{fig:lifecycle}
\end{figure}
} 

\hide{

The high-level functional components of the current
Alexandria system are shown in Figure \ref{fig:Alex-overall-flow}.

The domain scoping interface is illustrated in
Figures \ref{fig:scoping-02-terms-and-multiple-open-concepts}
and
\ref{fig:scoping-05-terms-concepts-indicators}.

} 

\section{System Archtecture}
\label{sec:arch}

\begin{figure} 
\vspace*{-2mm}
\centerline{\includegraphics[width=3.5in]{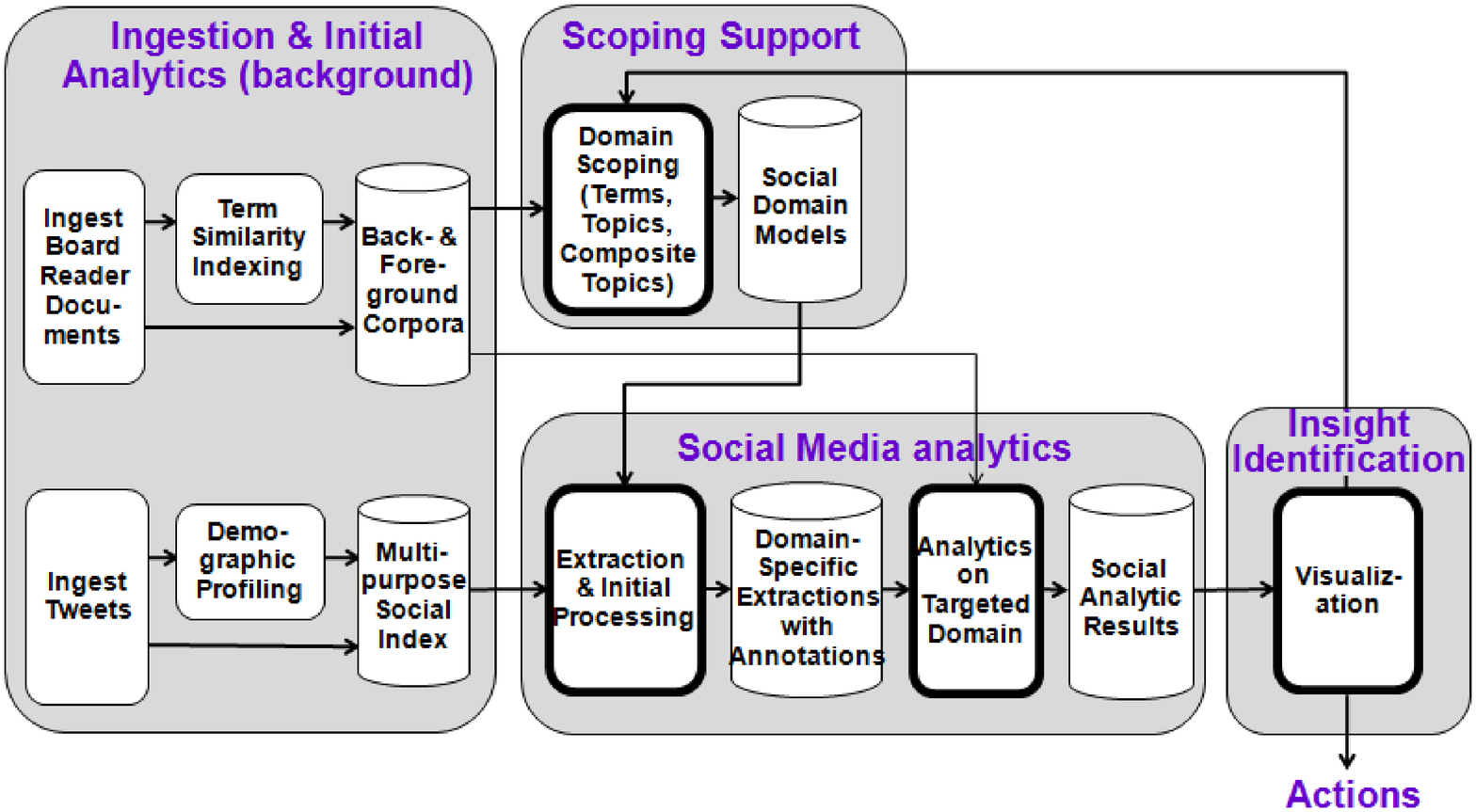}} 
\vspace*{-3mm}
\caption{Alexandria social media exploration process: 
         Ingestion and initial analytics in the background;
         Domain Modeling using text analytics;
         a broad variety of Social Media Analytics;
         and identification of actionable insights through
         visualizations.  Insights can lead to iterative modifications
         of the Domain Model and application of further analytics.}
\label{fig:arch-flow}
\vspace*{-2mm}
\end{figure}

\begin{figure*}[t]
\vspace*{-2mm}
\centerline{\includegraphics[width=5.5in]{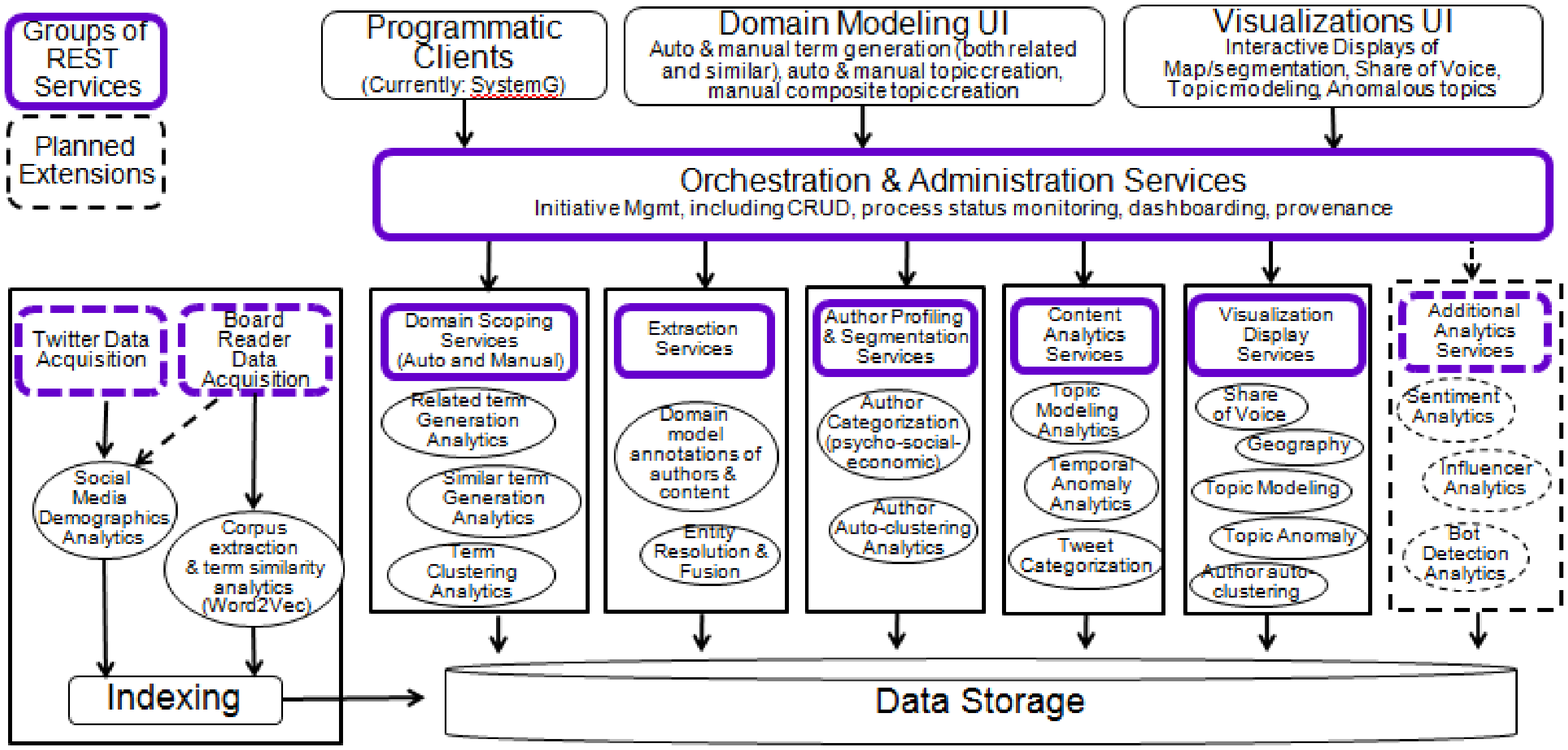}} 
\vspace*{-3mm}
\caption{Alexandria supports loosely coupled 
         RESTful services that orchestrate and invoke many
         functionalities, all sharing a common data store}
\label{fig:arch-REST}
\vspace*{-3mm}
\end{figure*}

The Alexandria architecture will be described
from three perspectives:
(a) the overall processing flow
(see Figure \ref{fig:arch-flow}), 
(b) the families of REST APIs supported (Figure \ref{fig:arch-REST}),
and
(c) the key systems components. 
These descriptions will include discussion of how
the architectural choices support the long-term and short-term
goals.

Overall, the current Alexandria architecture flow 
shown in Figure \ref{fig:arch-flow} expands on 
Figure \ref{fig:Alex-iterative-usage},
and is focused
on supporting rapid exploration, analytics processing, and
visualization of Twitter data in a collaborative
environment, that is, on parts of goals \LGtwo, \LGthree\ and \LGfour,
and all of goals \SGone\ through \SGsix.
There are two forms of background processing.
One is to ingest and index the Tweets, and also includes
author-by author processing of tweets to extract demographic
attributes, such as gender, geographic location, 
and one to ingest, process, and index background
text corpora.  (This demographic processing uses the IBM Research SMARC
system \cite{SMARC-in-IEEE-Big-Data-2013}, a precursor to IBM's Social Media Analytics product
\cite{SMA},
but other systems could be used).
The results are placed into a LUCENE SOLR main-memory index to
enable rapid searching, including against the Tweet text bodies,
a key enabler for goals \LGfour, \SGone, \SGthree, and \SGfive.
The other background processing is to ingest, process, and index 
various background corpora to support text analytics.
As described in further detail in Section \ref{sec:scoping}
below,
this is used to support the interactive domain scoping activity,
relevant to goals \LGtwo, \SGone, \SGtwo.
And as describe in Section \ref{sec:analytics},
this is also used
to support the anomalous topics analytics and view (goals \LGtwo, \SGthree).

Referring again to Figure \ref{fig:arch-flow},
once a Domain Model is established for a Project,
the Social Media analytics processing is performed.
This is described in more detail in section \ref{sec:analytics} below.
\hide{
One stage of this is to extract all of the relevant Tweets
and author profiles from the background, perform
some transformations and annotations on the data and
store the results, all to support a variety of analtyics.
At present the annotated data is stored in CouchDB, but
could also be placed onto, e.g., SPARK.
} 
After extraction and annotation,
the desired analytics are invoked through REST APIs by
an orchestration layer and the results are again
placed into CouchDB.
Finally, these can be accessed through several interactive
visualizations.

\hide{

\begin{figure} 
\vspace*{-2mm}
\centerline{\includegraphics[width=3.5in]{arch-04-system-components.eps}} 
\vspace*{-3mm}
\caption{System components that support the Alexandria capabilities.
(Words in gray indicate planned extensions.)} 
\vspace*{-2mm}
\label{fig:arch-04-system-components}
\end{figure}

} 

As illustrated in Figure \ref{fig:arch-REST},
most capabilities in Alexandria are accessed through REST services,
which is the basic approach to supporting goals \LGone, \SGthree\ and \SGsix.
\hide{
The current system is somewhat limited 
with regards to support for exploration and collaboration
(\LGthree, \SGfour) because the orchestration of components is
mostly hard-wired; but we are currently using the REST APIs
to develop a much more flexible orchestration and dashboarding layer.
}
For capabilities involving large data volumes, the 
data is passed ``by reference'' for increased performance (\LGfour, \SGfive).
At present the REST services are grouped more-or-less according
to the architectural flow of Figure \ref{fig:arch-flow}.
(It is planned to REST-enable the background processing.)
The REST services rely on a shared logical Data Store, which
is currently comprised of LUCENE and CouchDB.  
This can be extended to other storage and access technologies
without impacting the REST interfaces (goals \LGone, \LGfour, \SGfive).

The REST-based architecture has already been applied to 
enable a rapid integration of Alexandria capabilities with
IBM Research's SystemG \cite{system-g-home-page},
a graph-based system that also supports social media analytics.
In particular, the Alexandria Domain Models are now
accessible to SystemG services, and the SystemG UI has been 
extended to support both Domain Scoping and 
Alexandria
analytics views.
 

Alexandria exists as a software layer that can access
raw repositories and streams of social media (and other) data,
and that resides on top of several application, middleware, and
data storage technologies.
The system currently uses the GNIP Twitter reader 
and Board reader to access social media and web-accessible
data.  
The application stack is currently based on LUCENE, CouchDB, and HDFS
for data storage and access,
Hadoop for cluster management, IBM's Big Insights, SPSS, and
Social Media Analysis for analytics, and finally TomCat and
Node.js to provide application server middleware.
Alexandria lives above these layers, and could be
extended to take advantage of other server capabilities
(goals \LGone, \LGfour, \SGthree, \SGfive).

\hide{
\begin{figure} 
\vspace*{-2mm}
\centerline{\includegraphics[width=3in]{arch-01-components.eps}} 
\vspace*{-3mm}
\caption{High-level architecture and key data repositories illustrating the 
        components and data flows involved when performing a 
        Social Media Investigation}
\label{fig:arch-components}
\end{figure}
} 

\hide{
One of the goals of this architecture is to automate the various steps
in the comprehensive social media investigation process, and to
curate, annotate and analyze end-to-end data as illustrated in Figure
\ref{fig:arch-REST} below. The steps include modeling and
discovery, extraction and annotation, analysis and visualization of
the investigation results, and most importantly the exploration
iterations as illustrated in Figure \ref{fig:arch-flow}.
} 



\hide{
The architecture consists of a set of RESTful services. Alexandria
Orchestration Services coordinates all the underlying services to
ensure that dependencies between models, extracted data, and analytics
en-gines are handled appropriately. Figure \ref{fig:arch-REST}
illustrates the services in-volved in the social media investigation
process, and also illustrates that additional analytics can easily be
plugged into the framework.

The extensibility of Alexandria has already been demonstrated
through some integration with System G \cite{system-g-home-page}, 
a broadly focused platform 
for social media analytics that uses rich graph-based analytics.
} 

\section{Domain Scoping}
\label{sec:scoping}

Domain Scoping addresses the challenge of constructing Domain Models.
A Domain Model
is typically represented as families of keywords and 
composite topics (a.k.a., text extractors),
which get applied to the corpus of text documents to realize the search
or filtering in the corpus. Traditionally, Domain Scoping is performed
by a subject matter expert who understands the domain very well and
can specify precisely what the particular queries and search criteria
should be for a given set of topics of interest. A central goal of Alexandria
is to simplify significantly the task of creation of Domain Models
as well as to lower the required domain expertise of the person 
creating Domain Models. To achieve that, we developed several
techniques that leverage text analysis and data mining in order to
assist at discovery and definition of relevant topics that will
drive creation of search queries. In particular, we describe our approach
for (1) discovery of relevant collocated terms, for (2) term
clustering, and for (3) similar term generation. 
As illustrated in Section \ref{sec:overview}, these three techniques combined together allow very easy,
iterative definition of terms and topics (i.e., sets of collocated terms)
relevant for a particular domain
with minimal input required from the user. 
Other scoping tools can be incorporated into Alexandria, e.g., 
a tool based on using an ontology such as DBPedia.

\subsection{Collocated Term Discovery }

Alexandria employs two techniques \textendash{} term frequency\textendash{}inverse
document frequency (TF-IDF) score and collocation \textendash{} to
discover significant relevant terms to a specific set of seed terms.
Simply put, what Alexandria does is find documents that seed terms
appeared within. This is called the \textquotedblleft{}foreground\textquotedblright{}
documents. It then harvests other terms that were mentioned in the
documents and computes their significance.

To support this analytic, we acquired sample documents \textendash{}documents
considered general and representative enough of many different topics
and domains \textendash{} as the \textquotedblleft{}background\textquotedblright{}
materials for this operation. For this purpose we collected a complete
week of documents (Sept 1-7 2014) from BoardReader. This extraction
amounts to about 9 millions documents. The documents were then indexed
in SOLR \cite{solr}, a fast indexing and querying engine based on
Lucene, for later fast access. Next we queried \textquotedblleft{}NY
Times\textquotedblright{} from this large set of documents, which
resulted in news articles in many different areas including politics,
sports, science and technology, business, etc. This set of documents
is used to build a dictionary of terms that are not limited to a specific
domain within a small 
sample. It is the basis for Alexandria
to calculate term frequency in general documents. 

From the foreground materials, Alexandria computes the significance
of other terms in the documents using TF-IDF scores. TF-IDF score
is a numerical statistic widely used in information retrieval and
text mining to indicate the importance of a term to a document \cite{manning99}.
The score of a term is proportional to the frequency of the term in
a document, but is offset by the frequency of the same term in general
documents. The TF-IDF score of a word is high if the term has high
frequency (in the given document) and a low frequency in the general
documents. In other words, if a term appears a lot in a document,
it may be worth special attention. However, if the term appears a lot in
other documents as well, then its significance is low. 

$$
\begin{array}{rcl}
TF-IDF &  =  & TF(t,d)\times IDF(t,D)
\\
IDF(t,D) & = & log\frac{N}{|t\in D|}
\end{array}
$$

A collocation is an expression consisting of two or more words that
corresponds to some conventional way of saying things. They include
noun phrases such as \textquotedblleft{}weapon of mass destruction\textquotedblright{},
phrasal verbs like \textquotedblleft{}make up\textquotedblright{}
and other stock phrases such as \textquotedblleft{}the rich and powerful.\textquotedblright{}
We applied collocation to bring in highly relevant terms as phrases
when the words collocate in the document and would make no sense as
individual terms. More details of this technique can be found in \cite{bdvj-nplm-03}.
Examples of these phrases are seen in Figure 1, for example, \textquotedblleft{}small
business,\textquotedblright{} \textquotedblleft{}retail categories,\textquotedblright{}
and \textquotedblleft{}men shirts.\textquotedblright{} 

For collocated term generation, the larger the corpus and the more accurate
the results will be. However a very large corpus will suffer from
efficiency and is not practical to use in an interactive environment
such as Alexandria. Our hypothesis is that a week of general documents
as a background corpus is a good enough representative of the bigger
corpus, but is small enough to calculate the TF-IDF and collocation
scores in a responsive manner. 

\subsection{Term Clustering and Similar Term Generation}

Alexandria uses a term-clustering algorithm based on semantic similarities
between terms to semantically group them into appropriate and strong
``topics''. Alexandria uses Neural Network Language Models (NNLMs)
that map words and bodies of text to latent vector spaces. Since they were initially proposed \cite{bdvj-nplm-03}, a great amount of progress has been
made in improving these models to capture many complex types of semantic
and syntactic relationships \cite{Mikolov-2013,PenningtonSM14}. NNLMs are generally trained in an unsupervised
manner over a large corpus (greater than 1 billion words) that contains
relevant information to downstream classification tasks. Popular classification
methods to extract powerful vector spaces from these corpora rely
on either maximizing the log-likelihood of a word, given its context
words \cite{Mikolov-2013} or directly training from the probabilistic
properties of word co-occurrences \cite{PenningtonSM14}. In
Alexandria, we train our NNLMs on either a large corpus of Tweets
from Twitter or a large corpus of news documents to reflect the linguistic
differences in the target domain the end user is trying to explore.
We also extended the basic NNLM architecture to include phrases that
are longer than those directly trained in the corpus by introducing
language compositionality into our model \cite{socher2013recursive,goller96,mikolov2010recurrent}.
This way, our NNLM models can map any length of text into the same
latent vector spaces for comparison. 

The similarity measure obtained to support the term clustering
is also used to generate new terms that are ``similar'' to the terms
already in a topic.

\hide{

One challenge we are facing with the Alexandria user experience design is
handling concepts that have been altered by the users. When concepts
are manually created, or automatically computed ones are enhanced during
iterations, the concept-clustering algorithm has to take that into
account. The concepts that users created or changed can be leveraged by
the clustering system, which will train a neural network, aiming to
learn a belief function of the users target task. For this computation
it is important to prevent over fitting, and techniques like dropout
\cite{Hinton2012} lead to better clustering results. These vector
space models can also be used to create intelligent interface features.
For example, we can leverage text snippet similarity to automatically
name a new term list that bares strong conceptual resemblance to a
list that was previously created. 

} 

\hide{

\subsection{Profile Extraction}

Once tweets are extracted from the SOLR index, the corresponding author\textquoteright{}s
profiles are compiled. Both the tweets and profiles are annotated
along the foci, and stored for the initiative in both
 CouchDB (noSQL
database) and SOLR indexes. 

Alexandria incrementally fetches from the Twitter decahose to maintain
a 6-month rolling window of tweets. We also incrementally perform
analytics to compile authors \textquotedblleft{}user\textquotedblright{}
profiles. Attributes such as locations (used in showing geographic
distribution), whether authors are parents, intend to travel, whether
they are business owner, are computed using tweets as evidence. The
analytics based on previous research work done at IBM \cite{SMARC-in-IEEE-Big-Data-2013}
has shown to show around 82 \textendash{} 94 \% accuracy. 

} 

\section{Analytics Views}
\label{sec:analytics}

This section briefly surveys two of the four main
analytics algorithms currently supported by Alexandria;
the others are omitted due lack of space.

\subsection{Profile Extraction}

As a pre-cursor to the other analytics in Alexandria, the tweets identified by
the composite topics are 
extracted from the SOLR index and the corresponding
authors' profiles are compiled.  Both the tweets and profiles are
annotated along the composite topics, and stored for the Project in both
CouchDB (noSQL database) and SOLR indexes.  Alexandria incrementally
fetches from the Twitter decahose to maintain a 6-month rolling
window of tweets.  We also incrementally perform analytics to compile
authors' “user” profiles.  Attributes such as locations (used in
showing geographic distribution), whether authors are parents, and intent
to travel, 
are computed using tweets
as evidence.  The analytics based on previous research work done at
IBM \cite{SMARC-in-IEEE-Big-Data-2013}
has shown to show around 82\% to 94\% accuracy.

\hide{

\begin{figure}
\begin{small}
\begin{center}
\begin{tabu} to 3.4in { | X[c] | X[c] | X[c] | }
\hline
Stand-alone machine & 10 nodes with 80 mappers & 10 nodes with 17K mappers  \\
\hline 
$\sim$15 hours & $\sim$2 hours & $\sim$1 hour \\
\hline
\end{tabu}
\end{center}
\end{small}
\caption{Time used for ingestion and indexing of
         1 month of English-language
         tweets from Twitter Decahose ($\sim$400 million tweets, with 
         resulting index $\sim$128 GB)}
\label{fig:tweet-ingestion-times}
\end{figure}

\begin{figure}
\begin{center}
\begin{small}
\begin{tabu} to 3.4in { | X[c] | X[c] | }
\hline
Number of tweets & Time for extraction and DB writes \\
\hline 
$\sim$0.5M (453,931) & 4 min 29 sec \\
\hline
$\sim$1M (939,241) & 11 min 31 sec \\
\hline
\end{tabu}
\end{small}
\end{center}
\caption{Time used for extraction, annotation, and storage
         of tweets matching a Domain Model}
\label{fig:tweet-extraction-times}
\end{figure}

} 

We provide a brief illustration of the running time of various steps.
The current system is focused on a fixed set of English-language
Tweets from the Twitter Decahose (10\% of all Tweets).
With regards to background ingestion and initial processing,
the current Alexandria infrastructure uses a 4 node cluster,
with 1 as master and 3 as slaves; each node has
64MB of memory.
We focus on the time needed to process through Alexandria.
If a serialized machine 
were to be used then the extraction would be  about 15 hours; 
With 10 nodes and 80 mappers there is a stong time reduction
down to about 2 hours.  Increasing to 17K mappers (the maximum number)
brings the time to about 1 hour.

\hide{

The parallelism afforded by the 4 nodes
gives a substantial increase.
The table shows the time needed to ingest and index one day's worth
of the English-language tweets from the Twitter Decahose
(about 400 million tweets), to build an index 
(with size about 128 GB).
While the use of 80 mapper services reduces the time needed
to about 2 hours per day, using 17K mappers (the maximum available)
reduces the time to about 1 hour.  

} 

We also measured the end-to-end clock time for performing
the extraction and annotatoin stage for a set of tweets.
With a corpuus of almost half a million tweets (452,201) the elapsed time was
4 minutes 29 seconds.
With a corpus of almost a million tweets (949,241) it took 11 minutes and 31 seconds.
(The numbers are not linear probably because the system is
running on cloud-hosted virtual servers, which are subject
to outside work loads at arbitrary times.)
The processing includes writing the formated data into both
a CouchDB and a SOLR database.
\hide{

In the current system there are four parts to the extraction
processing step.
First, SOLR is used to create an index in main memory specific
to the targeted tweets.
Second, the actual tweet data is pulled into SOLR.
Third, the tweet data is pulled over the wire into
the extraction service component.
Finally, this service annotates the tweet and author information,
and writes them over the wire into three databases
(two couchDB and one SOLR). 

} 
Looking forward, we expect to move towards an architecture 
with a single indexed data store, so that
we can perform the annotations ``in-place''.

\subsection{Temporal Anomoly}

Lastly, Alexandria performs topic analytics to help the user explore
the topics discussed among tweets.  Unlike many available topic
detection algorithms \cite{twitter-emerging-trends-2011},
we define anomalous topics as terms that
suddenly receive attention in a specific week when compared to the
rest of the weeks in the data set.  Alexandria uses a technique
similar to the event detection domain \cite{beyond-twitter-trending-2011}.
It extracts terms from
tweets, compute TF-IDF scores and frequencies and only retain terms
with high TF-IDF score and high frequency.  To calculate anomaly score
for a term, we consider the frequency of the term in each week and its
frequency over all the weeks in the data set. If the term's frequency
and score deviate a lot in a particular week from what it normally has
over all, the term is considered anomalous.  There could be an event
or and emerging trend that caused the buzz, and hence people discuss more
about the term in that week. This can trigger the user to look further
to correlate research on events in that week.  Following shows the
formulas used for the calculation.  

\begin{small}
$$
\begin{array}{rcl}
{\rm anomalyScore}(term_i, week_j) & = &
  \frac{{\rm normFreq}(term_i, week_j)}{{\rm normFreq}(term_i,{\rm all\_weeks})}
\\
{\rm normFreq}(term_i, week_j) & = & 
  \frac{{\rm count}(term_i, week_j)}{{\rm maxCount}(week_j)}
\\
{\rm normFreq}(term_i, {\rm all\_weeks}) & = &
  \frac{{\rm count}(term_i, {\rm all\_weeks})}{{\rm maxCount}({\rm all\_weeks})}
\end{array}
$$
\end{small}

\section{Meta-data Support for Iterative Exploration}
\label{sec:meta-data}

Alexandria has been designed to support rapid, iterative, collaborative exploration of 
a domain including the usage of multiple analytics (goals \LGthree, \SGfour, \SGsix).
This is enabled in part by the disciplined use of REST APIs to wrap
the broad array of analytics capabiliites (see Figure \ref{fig:arch-REST}).
But the fundamental enabler is the strongly data-centric approach taken
for managing the several Projects that are typically created 
during the investigation of a subject area.

Data about all aspects of a Project (and pointers to more detailed information)
is maintained in a CouchDB document, called {\em ProjectDoc};
this can be used to support a dashboard about project status, and to 
enable invocation of various services.
For example, the ProjectDoc holds a materialized copy of the domain model
used to select the tweets and authors that are targeted by the Project.
It maintains a record of which analytics have been invoked,
and also maintains status during the analytics execution to enable
a dashboard to show status and expected completion time to the end-user.
Provenance data is also stored, to enable a determination of how
data, analytics results, and visualizations were created in case
something needs to be reconstructed or verified.

The ProjectDoc provides a foundation for managing flexible, ad hoc styles of
iterative exploration.
For example, with the ProjectDoc it is easy to support ``cloning'' of
a Project to create a new one, and to combine the Topics and Composite Topics from
multiple Projects to create a new one.
It also allows for maintenance of information about whether analytics results
have become out-of-date, and to support the
incremental invocation of analytics,
e.g., as new tweets become available.
It also supports the inclusion of new Composite Topics into a Project's domain model,
along with controlled, incremental computation of the analytics for these additions. 

\section{Related Work}
\label{sec:related}

\hide{
A precursor to Alexandria is LARIAT \cite{LARIAT-SCC-2014}.

Reference \cite{analytics-process-mgmt:DAB-2014}
describes the analytics process lifecycle.
} 

Many papers focus on understanding social media. Various social media
studies provide understanding of how information is gathered. For
instance, \cite{hurricane-sandy-social-CHI-2014} analyses community
behaviors of social news site in the face of a disaster, \cite{twitter-H1N1-2010}
studies information sharing on Twitter during bird flu breakout, and
\cite{health-info-online-CHI-2014} studies how people use search
engines and twitter to gain insights on health information, providing
motivation for ad hoc exploration of social data. 
Fundamentally, the authors of \cite{rajaraman2011mining} elaborated on design features needed
in a tool for data exploration and analysis, and coined the term 
\textquotedblleft{}Information
Building Applications.\textquotedblright{} They emphasized the support
for tagging and categorizing raw data into categorization and the
ability to restructure categories as their users, students, understand
more about the data or discover new facts. The authors also emphasized
the necessity of supporting fluid shift between concrete (raw data)
and abstract (category of data) during the validation and iteration
process, especially when faced with suspicious outcomes. While the
paper discussed specifically about a tool for exploring streams of
images, the nature of the approach is very similar to the process
of exploring social media we are supporting in Alexandria. 
From another direction, as discussed in \cite{analytics-process-mgmt:DAB-2014},
an environment for analytics exploration, and application of the results,
must support rich flexibility for 
pro-active knowledge-workers, and incorporate best practice approaches
including Case Management and CRISP-DM \cite{CRISP-DM}
at a fundamental level.
Because project management in Alexandria is based on data-centric principles
(Section \ref{sec:meta-data}),
along with the services-API-centric design,
the system lays the foundation
for the next generation of support for the overall analytics lifecycle.

Another novelty in our work is the combination of
various text analytics and social media exploration tools
into a broad-based solution for rapid and iterative domain modeling.
While many tools
exist, such as Topsy \cite{topsy}, Solr \cite{solr}, Banana \cite{banana},
we discovered that these tools do not support well the process and
the human thoughts in gathering quality results. The existing tools
typically tend to aid in a fraction of the overall exploration task
needed. 
More comprehensive, commercial tools such as HelpSocial \cite{HelpSocial}
and IBM Social Media Analytics \cite{SMA} are geared towards a complete
solution. However, these tools require employing a team of consultants
with deep domain expertise to operate as consulting services. Their
support for the exploration process is not trivial and relies heavily
on human labor and expertise. 
In terms of the research literature, 
Alexandria is helping to close a key gap in research on tooling for
data exploration that was identified in \cite{BertiniL09}.


\hide{
A precursor to Alexandria is 
the LARIAT system described in \cite{LARIAT-SCC-2014}.
Reference \cite{analytics-process-mgmt:DAB-2014}
describes the analytics process lifecycle.
Finally, related
work for analytics methods is addressed in the previous sections. 

} 

\section{Conclusions and Directions}
\label{sec:conc}

This paper describes the Alexandria system, which provides 
a combination of features aimed at enabling business analysts
and subject matter experts to more easily explore and derive
actionable insights from
social media.
The key novelties in the system are:
(a) enabling iterative rapid domain scoping that takes advantage of
several advanced text analytics tools, and
(b) the development of a data-centric approach
to support the overall lifecycle 
of flexible, iterative
analytics exploration in the social media domain.

The Alexandria team is currently working on enhancements in
several dimensions.  
Optimizations are underway, including a shift to
SPARK for management and pre-processing of the background corpora
that support the rapid domain scoping.
Tools to enable comparisons between term generation strategies
and other scoping tools
are under development.
A framework to enable ``crowd-sourced''
evaluation and feedback about the accuracy of extractors
is planned.
The team is working to support multiple kinds
of documents (e.g., forums, customer reviews, and marketing content), 
for both background and foreground analytics.
The team is also developing a persistent catalog for
managing sets of topics and extractors; this will
be structured using a family of industry-specific
ontologies.

More fundamentally, a driving question is how to bring predictive
analytics into the
framework.
A goal is to provide intuitive mechanisms
to explore, view and compare the results of
numerous configurations of typical machine learning
algorithms (e.g., clustering, regression).
This appears to be crucial for enabling
business analysts (as opposed to data scientists)
to quickly discover one-off and on-going insights that can be applied 
to improve  business functions such as
marketing, customer support, and product planning.

\section*{Acknowledgements}

We would like to acknowledge other team members, Richard Goodwin,
Sweefen Goh and Chitra Venkatramani.  We also would like to acknowledge
our colleagues from the SystemG project \cite{system-g-home-page}, 
including in particular
Ching-Yung Lin, Danny Yeh, Jie Lu, Nan Cao, Jui-Hsin (Larry) Lai,
and Roddrik Sabbah. 



%


{\small


} 

%
%

\end{document}